\documentclass[sigconf,balance]{acmart}

\AtBeginDocument{%
  }

\copyrightyear{2026}
\acmYear{2026}
\setcopyright{cc}
\setcctype{by}
\acmConference[CHI EA '26]{Extended Abstracts of the 2026 CHI Conference on Human Factors in Computing Systems}{April 13--17, 2026}{Barcelona, Spain}
\acmBooktitle{Extended Abstracts of the 2026 CHI Conference on Human Factors in Computing Systems (CHI EA '26), April 13--17, 2026, Barcelona, Spain}
\acmDOI{10.1145/3772363.3798342}
\acmISBN{979-8-4007-2281-3/2026/04}

\usepackage[utf8]{inputenc} 
\usepackage{euscript}
\usepackage{array}
\usepackage{multirow}
\usepackage{algorithm2e} 
\usepackage{enumitem}
\usepackage{balance}

\newcommand{\scope}[1]{\textbf{\textcolor[rgb]{0.12, 0.47, 0.7}{#1}}}
\newcommand{\placement}[1]{\textbf{\textcolor[rgb]{0.2, 0.63, 0.17}{#1}}}

\newcommand{\scopeV}[1]{\textit{\textcolor[rgb]{0.12, 0.47, 0.7}{#1}}}
\newcommand{\placementV}[1]{\textit{\textcolor[rgb]{0.2, 0.63, 0.17}{#1}}}

\newcommand{\eg}[0]{\textit{e.g.}} 

\newcommand{\pval}[1]{\emph{p}}
\newcommand{\pvalcorr}[1]{\emph{p-corr}}

\newcommand{\feature}[1]{\textit{#1}}

\newcommand{\sysshort}{{\sf AiRWeb}}
\newcommand{\ficind}{Noa}

\newcommand{\htag}[1]{$<${\tt \textcolor[rgb]{0,0,1}{#1}}$>$}

\begin{teaserfigure}
\centering
\includegraphics[width=\textwidth]{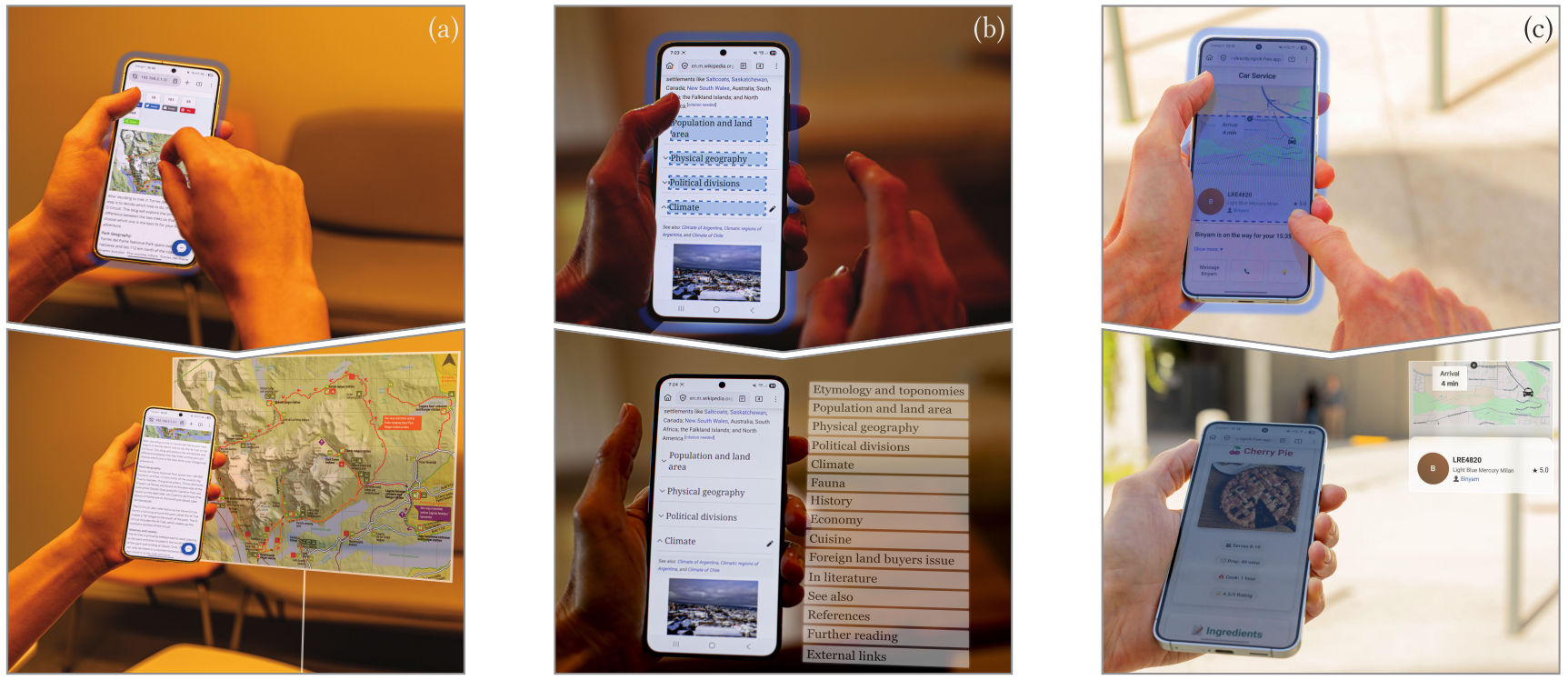}
\caption{Examples of Web page elements moved beyond the screen: (a) offloading a map above the coffee table so that it remains visible when scrolling down the page; (b) offloading all section headings of a document at once to get an interactive outline; (c)~offloading a rubber-band selection to monitor rendez-vous with a car pickup while reading about a cooking recipe.}
\label{fig:teaser}
\Description{Three panels, labeled (a), (b), and (c), each consist of a pair of photos. In all photos, the phone is held in the left hand. In the top photo of each panel, the right index finger selects an on-screen element; in the bottom photo, the element is persistently offloaded into augmented reality (AR) space. In panel (a), the top photo shows the right index finger flicking a hiking trail map, while the bottom photo shows the page scrolled and the map floating to the right of the phone. In panel (b), the top photo shows a downward drag forming a rubber‑band selection around car pickup details, and the bottom photo shows a cherry pie recipe on the phone while the selected car information remains in the field of view. In panel (c), the top photo shows all section headings of a long document selected, and the bottom photo shows those headings presented as a vertical outline to the right of the phone.}
\end{teaserfigure}

\begin{document}

\title{AiRWeb: Using AR to Extend Web Browsing Beyond Handheld Screens}

\author{Mengfei Gao}
\orcid{0009-0009-3372-1684}
\email{mengfei.gao@universite-paris-saclay.fr}
\affiliation{%
  \institution{Université Paris-Saclay, CNRS, Inria}
  \city{Gif-sur-Yvette}
  \country{France}
}
\author{Caroline Appert}
\orcid{0000-0002-3050-9284}
\email{caroline.appert@universite-paris-saclay.fr}
\affiliation{%
  \institution{Université Paris-Saclay, CNRS, Inria}
  \city{Gif-sur-Yvette}
  \country{France}
}
\author{Ludovic David}
\orcid{0000-0003-2341-9261 }
\email{ludovic.david@inria.fr}
\affiliation{%
  \institution{Université Paris-Saclay, CNRS, Inria}
  \city{Gif-sur-Yvette}
  \country{France}
}
\author{Emmanuel Pietriga}
\orcid{0000-0002-9762-0462}
\email{emmanuel.pietriga@inria.fr}
\affiliation{%
  \institution{Université Paris-Saclay, CNRS, Inria}
  \city{Gif-sur-Yvette}
  \country{France}
}

\renewcommand{\shortauthors}{Gao et al.}

\begin{abstract}
Browsing the Web on mobile devices is often cumbersome due to their limited screen space. We investigate a phone+AR Web browsing approach, AiRWeb, that leverages the structural properties of Web pages to allow users to seamlessly select and offload arbitrary Web content into the space surrounding them. Focusing on flexibility, AiRWeb lets users decide what to offload, when to do so, and how offloaded content is arranged, enabling personalized organization tailored to the task at hand. We developed a fully functional prototype using standard Web technologies, that covers the complete interaction workflow, from the selection of elements to offload from the phone to their manipulation in the air. Results from a preliminary study conducted using this prototype suggest that AiRWeb is learnable and usable, while also revealing open design challenges around offload mode activation in particular.
\end{abstract}


\maketitle

\section{Introduction}
\label{sec:intro}

Mobile Web browsing is constrained by the limited display area of handheld devices. As Augmented Reality (AR) technologies mature, handheld devices can be used in tandem with AR eyewear to increase effective screen real estate and extend interaction beyond the physical phone~\cite{Zhu:2020:BEB}. This combination can enhance the mobile Web browsing experience, which often involves navigating long pages and repeatedly jumping back and forth between dispersed pieces of content. Given that AR displays in mobile Web contexts are a novel interaction space, it is difficult to anticipate how users will appropriate them to organize content. Designing systems that enable flexible, user-driven offloading of Web elements provides an opportunity to support a wide range of individual strategies, without constraining users to predefined patterns.

\autoref{tab:positioning} provides an overview of key prior work on phone+AR displays, organized along two design dimensions:
\begin{itemize}
\item {\bf Offloading control} specifies who decides which content moves to the AR display. Some systems do not give end-users control over this: Beyond-the-Phone~\cite{Zhu:2025} relies on designer-defined application views that can be transitioned to AR; WebJump~\cite{Zeng:2023} requires developers to annotate webpage elements for offloading. In contrast, Push2AR~\cite{Wieland:2024} allows end-users to directly offload list elements using swipe gestures.
\item {\bf Offloading scope} captures the granularity of content that can be offloaded. Beyond-the-Phone~\cite{Zhu:2025} focuses on monolithic application views. Push2AR~\cite{Wieland:2024} supports offloading individual items from flat lists but does not support other types of Web elements.
\end{itemize}

\begin{table}[b]
\centering
\caption{Positioning of representative phone+AR systems along two dimensions: \emph{offloading control} (who determines what content is moved to AR) and \emph{offloading scope} (the granularity of offloaded content).}
\label{tab:positioning}
\begin{footnotesize}
\begin{tabular}{lcc}
\toprule
\textbf{System} & \textbf{Offloading control} & \textbf{Offloading scope} \\
\midrule
Push2AR~\cite{Wieland:2024} & User-driven & List items \\
Beyond-the-Phone~\cite{Zhu:2025} & Designer-defined & Monolithic application views \\
WebJump~\cite{Zeng:2023} & Developer-defined & Arbitrary Web elements \\
\textbf{\sysshort{}} & \textbf{User-driven} & \textbf{Arbitrary Web elements} \\
\bottomrule
\end{tabular}
\end{footnotesize}
\end{table}

Existing approaches limit users’ ability to decide what, when, and where to offload content from the phone to the air. To address this gap, we introduce \sysshort{}, a mobile Web browsing prototype that allows users to offload arbitrary parts of Web pages into AR and spatially arrange them around the user (\autoref{fig:teaser}). Unlike systems that rely on designer-defined views or developer-authored annotations, \sysshort{} operates on regular Web content, leveraging the generic DOM structure of Web pages to enable flexible, user-driven selection and offloading.

We contribute the first complete interaction design and implementation of a phone-based, AR-enhanced Web browsing system, from the selection of Web page elements to offload from the phone to their interactive manipulation in AR. We present the findings from a preliminary study performed using this prototype, that provides initial evidence about the relevance and usability of the approach, and also highlights open design challenges.

\section{Motivating Scenario}
\label{sec:mv}

\sysshort{} is designed to enable users to reorganize elements of arbitrary web pages in the space around their handheld device, supporting more flexible and efficient mobile web browsing. By allowing users to offload selected elements into AR, \sysshort{} helps reduce repetitive navigation actions and enables side-by-side access to relevant information, as we illustrate in the following short scenario.

While waiting for an appointment, \ficind{} starts planning for an upcoming hiking trip. They browse a Web page describing two alternative routes, that includes descriptions and a map showing both of them. To keep the map visible while the continue reading, \ficind{} activates offloading mode and selects \scopeV{the map alone}, placing a larger version \placementV{above the coffee table in front of them}  (Figure~\ref{fig:teaser}-a). They can then keep scrolling and reading while referring to the map at will.

Later, \ficind{} wants to learn more about Patagonia, the region in which they will be hiking. They go to Wikipedia on their phone. The page describing Patagonia is pretty long and features many sections. \ficind{} is just browsing serendipitously, curious to discover interesting things about the region. They offload \scopeV{all section headers at once} \placementV{near the phone} using a single gesture. They can now use the resulting document outline to jump between sections on the phone (Figure~\ref{fig:teaser}-b).

After their appointment, \ficind{} requests a ride home. While waiting for the car to arrive, they begin searching for a cherry pie recipe for guests arriving later that evening. As they want to keep track of the car’s location and estimated time of arrival, they perform a continuous selection of the \scopeV{relevant area} of the ride service page that shows the map and time informaton, and offload it into \placementV{a corner of their field of view} (Figure~\ref{fig:teaser}-c). This allows them to continue browsing recipes while monitoring the ride information without switching back and forth between Web pages.

This scenario illustrates the need for flexible \scope{selection at different granularities}: \scopeV{single elements}, \scopeV{groups of elements}, and \scopeV{arbitrary areas} of the page; and \placement{placement options}: \placementV{near the phone}, \placementV{fixed in the field of view}, or \placementV{anchored in the physical world}.

\section{\sysshort{}}

\begin{figure*}
\centering
\includegraphics[width=\textwidth]{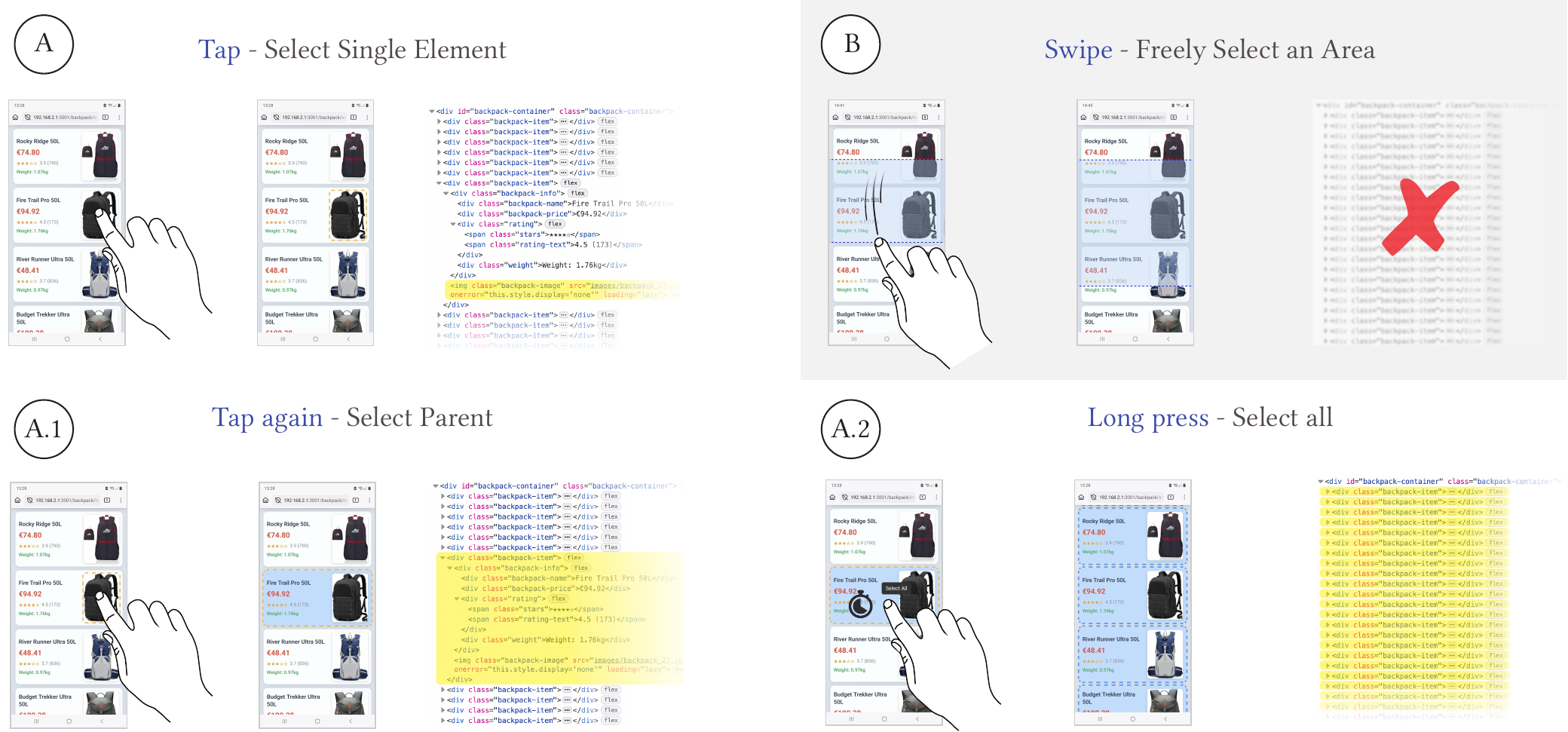}
\caption{Different types of element selections: (A)~Tapping on an element selects it; (B)~Performing a vertical drag makes a rubberband selection that can freely span elements regardless of the DOM hierarchy.; (A.1)~Tapping again on an already-selected element expands the selection to its parent in the DOM hierarchy; and (A.2)~Performing a long-press on an already-selected element pops up a button to expand the selection to all similar elements in the DOM hierarchy.
}
\label{fig:selections}
\Description{Four panels illustrate different types of element selections, showing mobile page interactions, resulting selection, and the corresponding DOM structure.}
\end{figure*}

We conducted a formative elicitation study (N=13) to investigate users’ expectations (UE) for offloading interactions. Participants proposed techniques for transferring elements from phone to AR, explored placement strategies, and described ways to discard offloaded content. From this study, we derived four key insights:

\begin{itemize}
\item \textbf{UE$_1$: Activation.} Offloading gestures typically started on or near the phone and ended in mid-air.
\item \textbf{UE$_2$: Selection.} Single-point selection was most common, with participants expecting the system to resolve ambiguities between nested or nearby elements.
\item \textbf{UE$_3$: Placement.} Anchoring of offloaded content was usually determined by the gesture’s endpoint.
\item \textbf{UE$_4$: Removal.} Participants used gestures such as throwing the element away or returning it to the phone.
\end{itemize}

{\bf Activation.} Participants used a regular mobile Web browser featuring the default, primarily touch-based, interactive capabilities. They were also asked to avoid conflicts with those capabilities when making proposals for offloading commands. Despite this, participants often proposed touch interactions that conflicted with those default capabilities.

To allow users to rely on simple gestures without conflicting with touch interactions, we introduced an explicit offloading quasimode. Users enter the mode by touching the side of the screen with the thumb holding the phone. A thumb silhouette appears in AR near the phone to indicate the action, and a blue halo signals that offloading mode is active, as illustrated in the top row of Figure~\ref{fig:teaser}. Those offloading quasimode exits automatically when lifting the thumb, returning the system to default browsing mode.

{\bf Selection.} With offloading mode active, content can be transferred via a \emph{finger-swipe offload} (touch+air) or an \emph{air-pinch offload} (air-only). During quick interactions, the system infers the target element (\textbf{UE$_2$}) by traversing the DOM to identify the first CSS block-level element at the gesture’s start point (e.g., a \htag{p} or a \htag{div} element).

For more granularity in selection control (\autoref{fig:selections}), users can \textbf{tap to select} the lowest-level element at that position, incrementally expanding the selection to parent elements; or they can long-press to select a collection of similar elements with a single action. Alternatively, users can \textbf{drag to select} an arbitrary rectangular region. During selection, \sysshort{} provides immediate visual feedback by modifying the CSS styling of selected DOM elements, clearly indicating what will be offloaded.

\begin{figure*}[t]
\centering
\includegraphics[width=.98\textwidth]{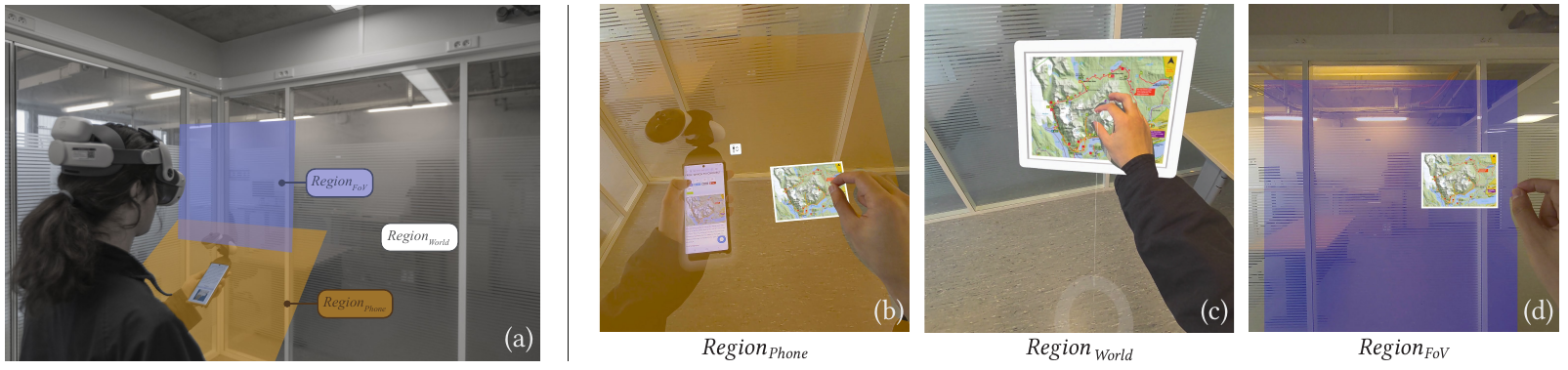}
\caption{\sysshort{} partitions space into three regions: $Region_{Phone}$ (orange), $Region_{FoV}$ (blue), and $Region_{World}$. Anchoring depends on the region where the element is released. Visual feedback guides users by revealing planar regions or highlighting the element’s relationship to physical surfaces.}
\label{fig:anchors}
\Description{Four photos show a user holding a phone with AR overlays indicating $Region_{Phone}$, $Region_{FoV}$, and $Region_{World}$. Subsequent photos illustrate interactions: the hand pinching a map in each region, visual feedback including planes, frames, and rays to surfaces.}
\end{figure*}

{\bf Management (Placement and Removal).} When using a \emph{finger-swipe offload}, the selection is by default offloaded near the phone and remains anchored to it. During an \emph{air-pinch} offload, the final anchoring depends on the gesture’s endpoint (\textbf{UE$_3$}). As shown in~\autoref{fig:anchors}, the hand typically traverses $Region_{Phone}$ first. An orange plane aligned with the phone is shown, indicating that the element being offloaded will be anchored to the phone if released now. Otherwise, if continuing to $Region_{World}$, \sysshort{} provides feedback via a vertical line anchoring the element to the nearest horizontal surface (\eg{}, the floor or a table), decorating the element with a 3D frame to suggest its anchroing into the physical environment. If continuing further, bring the element closer to the user's face, the hand enters $Region_{FoV}$. The vertical anchoring line and 3D frame disappear, and a blue plane signals fixed anchoring in the user's field of view. 

Elements anchored to the phone remain fixed but can be repositioned via pinch-and-drag. They can be toggled to a scrollable layout reflecting their original order. Offloaded items remain linked to the phone, acting as bookmarks and shortcuts: tapping an offloaded item scrolls the mobile page to show its counterpart, similar to what Push2AR does for list items~\cite{Wieland:2024}. Users can discard offloaded elements by throwing them away (\textbf{UE$_4$}).

\section{Implementation}
\label{sec:impl}

\sysshort{} is a fully functional prototype built entirely with Web technologies. On the smartphone, it runs as a Manifest V3 browser extension implemented in JavaScript. One injected script handles core functionalities such as gesture detection, element selection, and feedforward for offloading. A second background script renders selected elements to offscreen buffers, ensuring visual consistency with the original page. This is important because the mobile browser and the WebXR client on the AR headset differ in screen dimensions, aspect ratios, fonts, and rendering behavior, which would necessarily introduce appearance differences if the renderings were performed independently. Instead, elements are captured on the phone and pre-transmitted before the offload gesture, reducing latency and preserving fidelity.

On the AR headset, a WebXR client built with A-Frame and Three.js receives the offloaded elements, tracks users’ hands for mid-air interactions, and monitors the smartphone’s position to anchor elements in $Region_{Phone}$. For precise, low-latency tracking, a motion controller is attached to the smartphone, providing performance comparable to external tracking systems without requiring a fixed setup. A NodeJS server handles communication between devices using Socket.IO. The system has been evaluated on a Meta Quest 3 headset and a Samsung Galaxy S10 running Firefox nightly.

\section{Preliminary Evaluation}
\label{sec:fb}

We conducted a user study (N=12, aged 22--37) with researchers in HCI/Visualization to get preliminary feedback about \sysshort{}’s usability and practical utility. The study comprised a \textit{discovery phase}, where participants explored ten features to assess discoverability; and a \textit{scenarios phase}, where they completed five contextualized tasks adapted from our motivating scenarios (Section~\ref{sec:mv}).

\begin{figure*}
    \centering
    \includegraphics[trim={11cm 0 0 0},clip,width=0.95\textwidth]{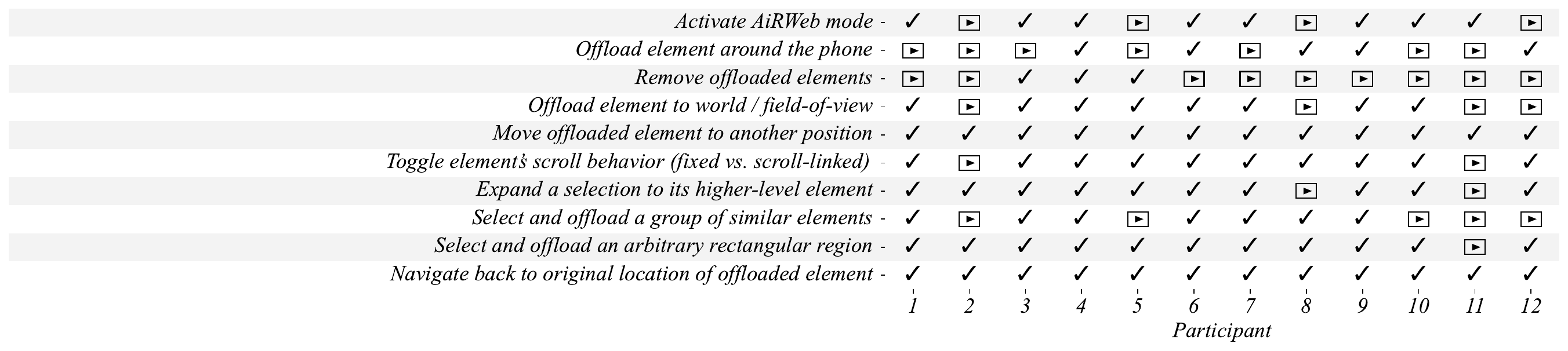} \\
\vspace{-1em}
    \caption{Help video consultations by feature $\times$ participant. Each cell indicates whether a participant (column) consulted the help video for the feature (row): \checkmark = no video consulted, \fbox{$\blacktriangleright$} = video consulted.}
    \label{fig:disc-matrix}
    \Description{Discoverability matrix of help video consultations across features and participants. Rows represent features of the interface, and columns represent individual participants. Cells indicate whether a participant consulted a help video for a given feature: a checkmark (✓) indicates no video consulted, while a play icon represents that the participant accessed the help video.}
\end{figure*}

{\em Discovery Phase.} Participants explored each feature at their own pace, consulting a help video if needed. Average exploration time was 32 minutes (median 23 minutes) for all ten features, always presented in the same order, covering the full interaction workflow and selection/placement options. As shown in~\autoref{fig:disc-matrix}, participants sought more help by watching videos in two cases: to offload elements, and to discard them (2nd and 3rd rows, respectively). This was due to misunderstandings about quasimode activation and discard gestures. Participants did not seek as much help about subsequent features, as they learnt from prior exploration steps.

{\em Scenarios Phase.} Observations in this phase suggest an effective and personalized appropriation of \sysshort{}. All participants successfully completed the 5-step scenario. We also observed variations in how participants selected and offloaded elements. Eight participants consistently used pinch gestures to offload a selection, while two participants always relied on flick gestures. When scenarios required selecting nested elements (\eg{}, A.2 in Figure~\ref{fig:selections}), two-thirds of participants used multiple taps (\feature{expand selection to enclosing elements}), whereas one-third preferred using drag gestures to select a rectangular area that enclosed the target element (\feature{select an arbitrary rectangular region}).

{\em User Experience.} Participants completed the \textit{User Experience Questionnaire (UEQ)}\footnote{\url{https://www.ueq-online.org}} at the end of the study. Ratings were positive across all dimensions: Attractiveness (M=2.13), Stimulation (M=2.27), Novelty (M=2.04), Efficiency (M=1.96) were \textit{excellent}, reflecting engagement and usefulness. Perspicuity (M=1.38) and Dependability (M=1.48) were above average, with some challenges reported for discard gestures and quasimode activation.

\section{Discussion}
\label{sec:disc}

Prior work has explored distributing Web interfaces across physical devices to overcome problems of limited screen space, either through developer-oriented frameworks~\cite{Yang:2014,Nebeling:2014,Klokmose:2015} or systems that allow users to redistribute elements across multiple displays at runtime~\cite{Nebeling:2016,Nebeling:2017,Park:2024}. These approaches typically focus on specific applications, predefined device ecologies, or semi-automatic adaptations, often in collaborative or task-specific settings.

\sysshort{} targets single-user, mobile Web browsing activities and investigates how Augmented Reality can enhance them by enabling users to offload selected page elements around their smartphone. It emphasizes user-driven, on-the-fly control over selection and placement of arbitrary elements from rich, structured Web pages without requiring application-specific support, predefined layouts, or adaptation rules. Following a phone-centric approach~\cite{Zhu:2020:BEB}, \sysshort{} uses AR to extend the smartphone rather than replace it. Our fully-functional prototype, built entirely on existing Web technologies, demonstrates that carefully designed interaction can make user-driven spatial offloading of live Web content feasible today. As AR eyewear continues to miniaturize and improve, such approaches could become increasingly integrated into everyday mobile browsing activities, once the following challenges are addressed.

{\em Interaction Design Challenges.} The study highlighted several interaction design challenges. First, participants sometimes performed the discard gesture in the opposite direction than intended, reflecting ambiguity between two metaphors: either ``throwing away'' or ``putting back inside the phone''. We plan to explicitly support both. Second, visual feedforward for the offloading quasimode was misinterpreted by some participants as an instruction to press a physical button on the phone, underscoring the difficulty of conveying subtle, one-handed gestures across devices. We are exploring alternative quasimode activations, including dedicated hardware buttons or AR-mounted virtual buttons with vision-based detection of thumb contact, which will become more feasible as hand-object interaction tracking improves.

{\em System and Architectural Challenges.} Beyond questions about interaction design, the implementation of \sysshort{} revealed system-level challenges. Its emphasis on high-fidelity visual consistency between phone and AR renderings limits offloaded elements to static snapshots for now. Supporting dynamic content while maintaining visual consistency would likely require streaming approaches (\eg{}, via WebRTC), raising issues of performance, latency, and network bandwidth overuse. Similarly, interaction with offloaded content is not yet supported, highlighting the need for tighter coupling between input, state, and execution contexts across mobile browsers and AR environments.
We are also working toward updating offloaded elements from background tabs. Combined with interactive offloaded content, this would better support parallel Web browsing, a task that is particularly challenging on mobile devices~\cite{Hasan:2017}.

\balance
    
\bibliographystyle{ACM-Reference-Format}
\bibliography{main}

\end{document}